\documentclass[11pt]{amsart}

\newtheorem{theorem}{Theorem}

\begin{document}

\title{Reducing the Space Used by the Sieve of Eratosthenes
  When Factoring}

\author{Samuel Hartman and Jonathan P.~Sorenson}
\address{Computer Science and Software Engineering,  Butler University,
  Indianapolis, IN, 46208 USA}
\email{jsorenso@butler.edu}

\begin{abstract}
  We present a version of the sieve of Eratosthenes that can factor
  all integers $\le x$ in $O(x \log\log x)$ arithmetic operations
  using at most $O(\sqrt{x}/\log\log x)$ bits of space.
  This is an improved space bound under the condition that the
    algorithm takes at most $O(x\log\log x)$ time.
  We also show our algorithm performs well in practice.
\end{abstract}

\date{\today}

\maketitle


\section{Introduction}\label{sec:intro}

Using the Sieve of Eratosthenes, we can factor all the integers up to
  a given bound $x$ using $O(x\log\log x)$ arithmetic operations.
This is optimal to within a constant factor;
  by the Erd\H{o}s-Kac theorem \cite{EK40}, the average number of prime
  divisors of an integer $n\le x$ is $\log\log x$.
Due to the work of Brent \cite{Brent73} and Bays and Hudson
  \cite{BH77} we know that the sieve of Eratosthenes can be
  \textit{segmented} so that it uses at most around 
  $\sqrt{x}$ space.
Note that we do not count the size of the algorithm output, as it might be
  streamed or consumed as it is generated.
Also, we measure time in arithmetic operations on $O(\log x)$-bit integers,
  and we measure space in bits used.
So $\sqrt{x}$ words of space would require $O(\sqrt{x}\log x)$ bits.
Previous work does not pin down exactly how many bits of space is used.

In this short paper, we present a new space complexity result for
  prime sieves that factor all integers up to a given bound $x$
  with optimal running time.
In particular,
  we show that all integers $n\le x$ can be completely factored
  in $O(x\log\log x)$ arithmetic operations using only
  $O( \sqrt{x}/\log\log x )$ bits of space. 
  an improvement of a factor asymptotic to $(\log x)\log\log x$ in space use,
  if we assume $\sqrt{x}\log x$ as the previous mark.
This savings in space is realized by applying two simple techniques:
  (1) packing the prime divisors of an integer $n\le x$ in $\log x$ bits,
  and (2) compressing the list of primes $p\le \sqrt{x}$ by storing
  only the gaps between primes, which requires an average of only $\log\log x$
  bits per gap.
  This second idea is due to Pritchard \cite{Pritchard83}.

\subsection*{Related Work}

Our result is only of interest if we focus on the special case of
  completely factoring all integers $\le x$ 
  using at most $O(x\log\log x)$ arithmetic operations.
If we allow only finding primes or allow our running time to creep up
  by even a logarithmic factor, $x^{1/3+\epsilon}$ space is possible.

The Atkin-Bernstein sieve \cite{AB2004} can find all primes $\le x$
  in $O(x/\log\log x)$ arithmetic operations using $O(\sqrt{x})$ space,
  and Galway \cite{Galway2000} showed how to reduce this space use to
  $x^{1/3+\epsilon}$ at the price of an increased running time to $O(x)$
  operations.
A segmented sieve of Eratosthenes using a wheel takes
  $O(x)$ operations and $O(\sqrt{x}(\log\log x)/\log x)$ bits of space
  \cite{Pritchard83} to find all primes up to $x$.
  See also \cite[Theorem 4]{Sorenson98}.

If we want complete factorizations but allow ourselves to exceed the
  optimal running time,
  the best algorithm to use, space-wise, is
  Helfgott's sieve \cite{Helfgott2020}; this algorithm uses
  $O(x \log x)$ arithmetic operations and only 
\\
  $O( x^{1/3} (\log x)^{2/3}\log\log x )$ words of space.

Both Galway's and Helfgott's space improvements do not adapt to
  working in an arithmetic progression.
Our algorithm will work to find factorizations in an arithmetic progression
  with no significant time or space penalty.
Indeed, the application to factoring integers of the form $p-1$
  for $p$ a prime in an arithmetic progression when computing
  primitive roots \cite{mcgown2023computation} 
  is the primary motivation for this work.

The rest of this paper is organized as follows.
In Section \ref{sec:thy} we present our algorithm, including descriptions of
  our two space-saving techniques, with a complexity analysis,
and in Section \ref{sec:code} we give some experimental results to
  show our algorithm is practical.

\section{Algorithm Description}\label{sec:thy}

We start with a segmented version of the sieve of Eratosthenes.
Let $\Delta$ be the size of our segment.  We choose the exact value later.
The integers up to $x$ are divided into $x/\Delta$ segments.
Let $x_1 < x_2 \le x$ be the lower and upper bounds on the current segment,
  so that $x_2=x_1+\Delta$.
Our goal is to factor all integers $n$ in the current segment, 
  with $x_1\le n < x_2$.

Let $S$ be an array of integer lists of length $\Delta$;
  $S[i]$ will hold the list of prime divisors of the integer $n=x_1+i$.
To begin, we make sure all of $S$'s lists are empty.
Let $P$ be the list of primes up to $\sqrt{x}$.
Although $\sqrt{x_2}$ is sufficient here, using $\sqrt{x}$ gives the same
  asymptotic complexity, and we will eventually need all those primes anyway.
\begin{quote}
\begin{verbatim}
S.clearall();
for(j=0; j<P.length(); j=j+1) 
{ 
  p=P[j]; 
  q0=x1+(p-(x1%p))%p; //** so p divides q0, q0>=x1
  for(q=q0; q<x2; q=q+p)
    S[q-x1].add(p); //** add p to q's divisor list  
}
\end{verbatim}
\end{quote}
The running time of this code is asymptotic to
$$
  \sum_{p\le\sqrt{x}} \left( 1 + \frac{\Delta}{p} \right)
  \quad= \quad
  \pi(\sqrt{x}) + \Delta\sum_{p\le\sqrt{x}} \frac{1}{p} \\
  \quad\sim\quad
   \frac{2\sqrt{x}}{\log x} + \Delta\log\log x,
$$
by the prime number theorem and Mertens's theorem \cite{HW}.
This routine is run once per segment.
With $x/\Delta$ segments, the total time, then, is
  $$O( x\sqrt{x}/(\Delta\log x) + x\log\log x).$$
To maintain an optimal running time we must have
$$ \Delta \gg \frac{\sqrt{x}}{(\log x)\log\log x}. $$
Since minimizing space implies minimizing $\Delta$, we set
  $\Delta$ to be a constant multiple of
$\sqrt{x}/((\log x)\log\log x) $.
We are now ready to discuss our two techniques.

\subsection{Packing Prime Divisors}

For a positive integer $n$, we have 
$$ \sum_{p\mid n} \log p \le \log n,$$
with equality if $n$ is squarefree.
This implies we can pack the prime divisors of an integer $n$ into
  $O(\log n)$ bits,
and our list of lists $S$ will require
  $O(\Delta\log x)$ bits total, or $O(\sqrt{x}/\log\log x)$ bits, meeting our
  promised space bound for $S$.

We use a fairly simple data structure that has two pieces for each
  integer $n$.
The first piece holds the variable-length binary representation of
  the primes found so far dividing $n$.
To make sure all the primes will fit, we don't store $2$, and we drop
  the least significant bit of all odd primes in the list.
Thus, to store $n=2\cdot 3\cdot 5\cdot 7\cdot 13=2730$, we don't store
  the $2$, and dropping the trailing $1$-bit gives us
  $3\rightarrow$\texttt{1},
  $5\rightarrow$\texttt{10},
  $7\rightarrow$\texttt{11}, and
  $13\rightarrow$\texttt{110}, giving the encoding 
\texttt{110 11 10 1} or
\texttt{11011101}
  (recall bit positions are numbered so position 0 is rightmost).
Note this requires 8 bits only, whereas $2730$ in binary is
  \texttt{101010101010}, using 12 bits.
In pseudocode presented below, the unsigned integer
  (64 bits in practice)
  holding these packed primes is called \texttt{list}.

Our second piece is an array indicating where the primes end.
For our example, the array would contain $[1,3,5,8]$.
Using bit shifts, masks, and bitwise operations, the array allows us
  to extract any prime in the list in constant time.
For example, to extract the prime from position $2$, 
  which is the third prime (we index arrays starting at 0),
  we know its bitlength is $5-3=2$, starting at position 3, so bit positions
  $3$ and $4$.
A shift and masking gives us the bits \texttt{11}.  Tacking on the missing
  \texttt{1} gives \texttt{111}, or $7$.

This second piece requires an array of pointers of length 
  $O(\log n/\log\log n)$,
  as this is the maximum number of prime divisors of $n$
  (we used $15$ in practice).
Each pointer stores a bit position bounded by $O(\log n)$, so
  $O(\log\log n)$ bits suffice (we used 8 bits in practice),
  for a total of $O(\log n)$ bits.
In the pseudocode below, this array is called \texttt{ptr[\,]},
  of length \texttt{plen}, indexed starting at zero.

The data structure supports constant time operations to add a new prime
  to the end of the list and to fetch any one prime by position (random access).

To add a prime \texttt{p} to the list:
\begin{quote}
\begin{verbatim}
pos=plen;    //** plen is the length of the ptr[] array
plen=plen+1;
if(pos>0) left=ptr[pos-1] else left=0;
pbits=(p>>1);                 //** strips trailing bit
list = list | (pbits<<left);    //** adds p on the end
ptr[pos]=left+bitlength(pbits); //** store where p ends
\end{verbatim}
\end{quote}
To get a prime from position \texttt{pos} in the list:
\begin{quote}
\begin{verbatim}
if(pos>0) left=ptr[pos-1] else left=0;
copy=(list >> left);  //** remove everything to the right
//** then mask to remove higher bits 
copy=(copy & ( (1<<(ptr[pos]-left) ) -1 ));
return (copy<<1)|1;   //** restore the trailing 1
\end{verbatim}
\end{quote}
Here we use the C/C++ style bit operations
left and right shift \texttt{<<}, \texttt{>>},
and bitwise and \texttt{\&} and or \texttt{|}.

\subsection{Compressing Small Primes}

The other major use of space comes from the need to store the list of primes 
  up to $\sqrt{x}$.
As each prime $p$ needs $O(\log p)$ bits, this is
$$ \sum_{p\le \sqrt{x}} \log p \sim \sqrt{x}$$
by the prime number theorem.
This is above our goal.

The idea for reducing this, due to Pritchard \cite{Pritchard83},
  is fairly simple.
Instead of storing each prime, we store the differences between
  consecutive primes.  
In the context of the algorithm, we only need to iterate through the primes
  in order, so adding the next gap to the current prime is quick and easy to do.
Since the average distance between primes below 
$\sqrt{x}$ is
$\frac{1}{2}\log x$, this requires only $O(\log\log x)$ bits, for a total
of $O(\sqrt{x}(\log\log x)/ \log x)$, well under our bound.
For example, instead of storing $2,3,5,7,11,13,17,19,\ldots$, we store the gaps
$1,2,2,4,2,4,2,\ldots$.
If we leave off the prime $2$, then all the gaps are even, so that we can
  store half the gaps, or $1,1,2,1,4,1,\ldots$.
In practice, the largest gap between primes does not exceed $500$ until
  over $300$ billion, meaning that half-gaps fit in 8-bit integers up
  to this point.

In theory, we have no guarantee that gaps will always be small.
However, the number of gaps that exceed, say, $(\log x)^3$, cannot exceed
  $\sqrt{x}/(\log x)^3$, since the sum of the gaps is $x$.
So in theory, to handle gaps larger than $(\log x)^3$, 
  we store the primes before and after
  each such large gap in a secondary list, 
  and simply put in $0$s in place of the large gap values 
    to tell us to look at this second list.

We have proved the following.
\begin{theorem}\label{maintheorem}
The Sieve of Eratosthenes can be improved to find the complete factorization
of all integers $n\le x$ using at most $O(x\log\log x)$ arithmetic
operations and $O(\sqrt{x}/\log\log x)$ bits of space.
\end{theorem}

As a side note, observe that the idea to store half-gaps can be generalized.
Let $m$ be a product of the first $k$ primes, where $k$ is small.
Store the primes in $\phi(m)$ separate lists by storing the gaps between
  them in each residue class.
Since the gaps will be multiples of $m$, we can store only the value of the
  gap divided by $m$.
But the number of primes in a residue class is asymptotic to 
  $x/(\phi(m)\log x)$ by Dirichlet's theorem, so the
  average gap is larger by a factor of $\phi(m)$.
This scheme, then, saves $\log_2 (m/\phi(m))$ bits per gap stored.
If we choose $k$ around, say, $\frac{1}{4}\log x$ so that
  $m$ is near $x^{1/4}$, then we save $O(\log\log\log x)$ bits,
  by Mertens's theorem.
This is negligible, but could pull the number of bits below,
  say, a power of $2$ or word boundary in practice, 
  making it worth considering in some special cases.

\section{Empirical Results}\label{sec:code}

\newcommand{\Plain}{\textsc{Plain}}
\newcommand{\Pack}{\textsc{Pack}}
\newcommand{\Gap}{\textsc{Gap}}
\newcommand{\Both}{\textsc{Both}}

We implemented four versions of the sieve of Eratosthenes:
  a version with no improvements (\Plain), 
  a version that uses our first idea to pack divisors (\Pack),
  a version that compresses the primes using gaps (\Gap),
  and a version that does both packing and uses prime gaps (\Both).
We coded these four algorithms in C++ and ran them on a linux PC.
The timing results appear in Table \ref{timestable}.

All of the algorithms factored all the integers in the interval
  $[10^{16}-10^{9}, 10^{16})$. 
To help verify correctness,
  they all reported the total number of primes found, $27147369$,
  and the total number of prime divisors found, $3883730055$
  (here $18=2\cdot 3^2$ counts as two prime divisors).
We varied the size of the segment, $\Delta$, for all four algorithms:
  $\Delta=2^{21}, 2^{23}, 2^{25}, 2^{27}$.
Note that 
  $2^{21}$ is about 2 million,
   a reasonable estimate for $\sqrt{x}/(\log x \log\log x)$ for
   $x=10^{16}$, and
  $2^{27}$ is just over $10^8=\sqrt{x}$.

The Plain and Gap versions allocate $\Delta$ arrays of 16 32-bit unsigned
  integers.
This is why we did not test near $x=10^{18}$; having enough RAM becomes
  a problem for $\Delta$ near $\sqrt{x}$.
The Pack and Both versions allocate $\Delta$ copies of 3 64-bit 
  unsigned integers, less than half the space.

We did not investigate using $\Delta$ linked lists to store divisors;
  we expect the allocation and deallocation of nodes would make it
  significantly slower than using a fixed-length array,
plus, storing a node pointer for each divisor uses up a lot of space.

The algorithm of Theorem \ref{maintheorem} is the \Both\ 
  algorithm using the smallest value for $\Delta$;
  the time for this algorithm choice is in boldface.

\begin{table}
\begin{tabular}{l|rrrr}
$\Delta$ & \Plain & \Pack & \Gap & \Both \\
\hline
$2^{21}=2097152$ & 	100.7 & 	91.2 & 	102.5 & 	\textbf{91.9}\\
$2^{23}=8388608$ & 	92.5 & 	74.6 & 	94.1 & 	75.0\\
$2^{25}=33554432$ & 	89.5 & 	66.0 & 	88.8 & 	65.8\\
$2^{27}=134217728$ & 	88.8 & 	61.5 & 	89.6 & 	62.5\\
\hline
\end{tabular}
\ \\

\caption{Running times in seconds
  on $[10^{16}-10^{9}, 10^{16})$
  for sieve of Eratosthenes variants}
\label{timestable}
\end{table}

These results were obtained on an Ubuntu linux desktop
using the \texttt{g++} Gnu compiler with level 2 optimization,
with the following hardware:
\begin{quote}
12th Gen Intel(R) Core(TM) i5-12500T \\
4.4 GHz Turbo, 2.0 GHz base CPU speed \\
18 MB cache, 64GB RAM
\end{quote}
We must caution the reader that these timing results are for a
  specific implementation and platform, and may not generalize.
That said, it seems clear that packing prime divisers not only
  saves space, but gives much better performance in practice,
  despite the extra number of operations.
Compressing the list of primes seems to have little impact on
  the running time.
We feel safe in recommending both improvements.

In conclusion,
the algorithm of Theorem \ref{maintheorem} minimizes space,
  and yet has a running time that is quite competitive with
  the plain version.

\section*{Acknowledgements}
Thanks to Ankur Gupta, Eleanor Waiss, and Jonathan Webster for 
  helpful discussions and useful feedback
  on an earlier draft of this paper.

The authors were supported in part by a grant from the
Holcomb Awards Committee.

The first author was an undergraduate student at the time
this work was done.

\nocite{Pritchard81,Pritchard83}

\bibliographystyle{plain}

\end{document}